\documentclass{elsart}
\usepackage{graphicx}
\usepackage{amsmath}
\usepackage{amssymb}
\begin{document}

\begin{frontmatter}

\title{Magnetic Field Effects on the Superconducting and Quantum Critical Properties
of Layered Systems with Dirac Electrons }

\author{E. C. Marino} and
\author{Lizardo H. C. M. Nunes}

\address{Instituto de F\'{\i}sica,  Universidade Federal do Rio de Janeiro,\\
Cx. P. 68528,  Rio de Janeiro-RJ 21941-972, Brazil}

\begin{abstract}

We study the effects of an external magnetic field on the
superconducting properties of a quasi-two-dimensional system of
Dirac electrons at an arbitrary temperature. An explicit expression
for the superconducting gap is obtained as a function of
temperature, magnetic field and coupling parameter ($\lambda_{\rm
R}$). From this, we extract the $B \times \lambda_{\rm R}$, $T\times
\lambda_{\rm R}$ and $B \times T$ phase diagrams. The last one shows
a linear decay of the critical field for small values thereof, which
is similar to the behavior observed experimentally in the copper
doped dichalcogenide $Cu_xTiSe_2$ and also in intercalated graphite.
The second one, presents a coupling dependent critical temperature
$T_c$ that resembles the one observed in high-$T_c$ cuprates in the
underdoped region and also in $Cu_xTiSe_2$. The first one, exhibits
a quantum phase transition connecting a normal and a superconducting
phase, occurring at a critical line that corresponds to a magnetic
field dependent critical coupling parameter. This should be observed
in planar materials containing Dirac electrons, such as
$Cu_xTiSe_2$.

\end{abstract}


\end{frontmatter}

\section{Introduction}\label{int}

A lot of attention has been devoted recently to
quasi-two-dimensional condensed matter systems presenting a band
structure such that the dispersion relation of the active electrons
corresponds to the one of a relativistic massless particle. The
kinematics of such electrons is described by a Dirac instead of a
Schr\"odinger term in the hamiltonian \cite{diracelec,diracelec1}.
This fact has a profound impact on the physical properties of the
system. The Fermi surface reduces to a point, the Fermi point, where
the density of states vanishes. This will drastically affect the
physical properties of such materials. In a previous work
\cite{npb1}, we investigated the superconducting properties of a
quasi-two-dimensional system of Dirac electrons and showed that they
are completely different from the ones presented by usual
Schr\"odinger electrons. There is, in particular, a quantum phase
transition connecting the normal and superconducting phases, which
is controlled by the magnitude of the effective superconducting
interaction coupling parameter. The Cooper theorem, therefore, is no
longer valid, as one should expect in the absence of a Fermi
surface.

Among the materials presenting Dirac electrons as their elementary
excitations, there are a few, which have been intensely focused
lately. These are high-Tc cuprates \cite{htc,htc1,htc2,htc3},
graphene \cite{graf,graf1,graf2}, carbon nanotubes \cite{carbnano}
and transition metal dichalcogenides \cite{dcalc,dcalc1}.

The interplay between superconductivity and magnetism is a subject
of central interest in any research involving superconducting
materials. In particular, a key issue, both from the basic and
applied physics points of view is the analysis of the effects of an
external magnetic field on the superconducting properties of a
system. In the present work, we study the effects of an applied
constant magnetic field, perpendicular to a quasi-two-dimensional
superconducting system containing Dirac electrons. As investigation
method, we use the effective potential for the superconducting order
parameter, which we evaluate both at $T=0$ and $T\neq 0$, as a
function of the applied field ($B$).

We firstly consider the zero temperature case and explicitly obtain
the superconducting gap as a function of the magnetic field and of
the renormalized effective physical coupling parameter
($\lambda_{\rm R}$) that controls the superconducting interaction.
This allows us to obtain the ($T=0$) ($B \times \lambda_{\rm R}$)
phase diagram, which presents a quantum critical line separating the
normal and superconducting phases. A renormalization group analysis
is then performed, demonstrating that the physical results do not
depend on the renormalization point.

We then study the nonzero temperature regime and explicitly
obtain the superconducting gap as a function of the temperature and
of the applied field. From this, we extract an implicit equation
relating the critical temperature with the critical magnetic field,
$T_c(T_c,B_c)$. This allows us to obtain the $T\times \lambda_{\rm R}$
phase diagram, which shows the critical temperature line as a
function of the coupling, for different values of the magnetic
field. This line resembles the one observed in the underdoped regime
of high-$T_c$ cuprates and also in transition metal dichalcogenides \cite{tmd}

Also from $T_c(T_c,B_c)$, we can derive the $B \times T$ phase
diagram, for different values of the coupling parameter
$\lambda_{\rm R}$. The critical line separating the normal and
superconducting phases, in this case, shows a linear decay of the
critical field, for small values of this field. This type of
behavior has been reported recently in the experimental study of the
copper doped transition metal dichalcogenide $Cu_x TiSe_2$
\cite{tmd} and also in intercalated graphite compounds
\cite{intgraf}. Since both systems potentially possess Dirac
electrons as elementary excitations, it is conceivable that the
peculiar properties of such electrons would be responsible for this
common behavior of the critical field. In connection to this point,
it would be extremely interesting to measure the critical magnetic
field as a function of doping in $Cu_x TiSe_2$, in order to compare
with the result derived from our $B \times \lambda_{\rm R}$ phase
diagram.

\section{Model}\label{model}

We investigate here the effect of applying a constant magnetic field
$B\hat{z}$ along the c-axis of a quasi-two-dimensional
superconducting electronic system containing two Dirac points.
Assuming that the active electrons correspond to these points, it
follows that the electron kinematics will be described by a Dirac
equation \cite{diracelec}. The electron creation operator will be
$\psi^\dag_{i,\sigma,a}$, where $i=1,2$ denotes the Dirac point and
$\sigma=\uparrow,\downarrow$, the z-component of the spin.
$a=1,...,N$ is an extra label, identifying the plane to which the
electron belongs. Alternatively, in a multi-band system, $a$ could
be used to specify the electron band.

The Lagrangian describing the system, in the presence of the
external magnetic field is given by
\begin{eqnarray}
\mathcal{ L }
 & = &
{\rm i} \ \overline\psi_{ \sigma a} \left[ \hbar \partial_{ 0 } +
v_{ F }\gamma^{i } \left( \hbar  \partial_{ i } + {\rm i}
\frac{e}{c} A_{ i } \right) \right] \psi_{ \sigma a} - \psi^{
\dagger }_{ \sigma a} \left( \mu_{ B }  \vec{ B } \cdot \vec{ \sigma
} \right) \psi^{ \dagger }_{ \sigma a}
\nonumber \\
& & +
 g
\left(
\psi^\dag_{1\uparrow a} \ \psi^\dag_{2\downarrow a}
+ \psi^\dag_{2\uparrow a} \ \psi^\dag_{1\downarrow a}
\right)
\left(
\psi_{2\downarrow b} \ \psi_{1\uparrow b}
+ \psi_{1\downarrow b} \ \psi_{2\uparrow b}
\right )
\,
\label{EqModel}
\end{eqnarray}

where $A_{ i }$ is the vector potential corresponding to $\vec B$,
$\vec \sigma$ are Pauli matrices and $\mu_B$ is the Bohr magneton.
The second and third terms, respectively, contain the coupling of
the magnetic field to orbital and spin degrees of freedom. As in
\cite{npb1}, we assume there is an effective superconducting
interaction whose origin will not influence the results of this
work. $g$ is the superconducting coupling constant, which is
supposed to depend on some external control parameter. In order to
make the lagrangian smooth, we define $g\equiv \lambda/N$. We use
the same convention for the Dirac matrices as in \cite{npb1}.

Performing a Hubbard-Stratonovich transformation we arrive at

\begin{equation}
\mathcal{ L } \left[ \Psi , \sigma \right]= - \frac{1}{ g }\
\sigma^{ * }  \sigma + \Psi_a^\dag \mathcal{ A } \Psi_a
\label{Eqsigmapsi}
\end{equation}
where
\begin{equation}
\mathcal{ A } =
\begin{pmatrix}
\tilde{ \partial_{ 0 } }  & - \tilde{ \partial_{ - } } & 0 & \sigma \\
- \tilde{ \partial_{ + } } & \tilde{ \partial_{ 0 } } & \sigma & 0 \\
0 & \sigma^{ * } & \tilde{ \partial_{ 0 } } & \tilde{ \partial_{ + } } \\
\sigma^{ * } & 0 & \tilde{ \partial_{ - } } &  \tilde{ \partial_{ 0
} } ,
\end{pmatrix}
\label{EqA}
\end{equation}

with $  \tilde{ \partial_{ 0 } } \equiv {\rm i } \ \left( \hbar
\partial_{0 } + \mu_{ B } B \right) $,
$  \tilde{ \partial_{ \pm } } \equiv {\rm i } \ v_{ F } \left( \hbar
\partial_{ \pm } + {\rm i } (e/c) A_{ \pm } \right) $ and
$ \partial_{ \pm } = \partial_{ 2 } \pm {\rm i } \ \partial_{ 1 } $.
The fermions are in the form of a Nambu field $ \Psi^\dag_a =
(\psi^\dag_{1\uparrow a}\ \psi^\dag_{2\uparrow a}\
\psi^\dag_{1\downarrow a}\ \psi^\dag_{2\downarrow a} ) $ and the
auxiliary Hubbard-Stratonovitch field  $\sigma$ satisfies the
equation,
\begin{equation}
\sigma = - g\ \left (\psi_{2\downarrow a} \ \psi_{1\uparrow a} +
\psi_{1\downarrow a}\ \psi_{2\uparrow a} \right ). \label{eqsigma}
\end{equation}
This shows that $\sigma^\dagger$ is a Cooper pair creation operator.



Integrating on the fermion fields, we obtain the effective action
for $\sigma$, namely
\begin{equation}
S_{\rm eff}\left( |\sigma| , B \right) = \int d^3 x \left( -\frac{ N
 }{ \lambda }|\sigma |^{ 2 }\right ) - {\rm i}N  \ln {\rm Det}\left[ \frac{
\mathcal{A} \left[ \sigma, B \right] }{
  \mathcal{A}  \left[ \sigma= 0, B = 0 \right]}\right ] \, \label{EqSeff}
\end{equation}

\section{ Effect of a magnetic field on the superconducting quantum phase transition at T = 0 }\label{T0}

The effective potential per $N$ corresponding to (\ref{EqSeff}) may
be obtained by a saddle point procedure, in which the field
configurations assume their ground state average values that
correspond to the classic lowest energy configurations.

Observing that the determinant of the matrix $\mathcal{A}$ is given
by
\begin{equation}
 \det \mathcal{ A } = \left[\tilde{ \partial_{ 0 } }^{ 2 } -
\tilde{
\partial_{ + } }\tilde{ \partial_{ - } } + | \sigma |^{ 2 }\right]^2 \,
\label{EqdetA}
\end{equation}

and choosing the asymmetric gauge, in which $ \vec{ A } = B ( 0, x )
$, we infer from the effective action (\ref{EqSeff}) that the
effective potential will be

\begin{equation}
V_{\rm eff}\left( |\sigma| , B \right) = \frac{ | \sigma |^{ 2 } }{
\lambda } -  \int \frac{ d^{2 } k  }{ \left(  2 \pi \right)^{ 2 } }
\int \frac{ d \omega  }{ 2 \pi } \ln \left\{ \frac{  \det
\mathcal{A} \left[ \sigma, B \right] }{
 \det \mathcal{A}  \left[ \sigma= 0, B = 0 \right]
} \right\} \, ,\label{EqVeff}
\end{equation}

where (\ref{EqdetA}), in momentum-frequency space, is given by
\begin{equation}
  \det \mathcal{A} \left[ \sigma, B \right]
 = \left \{
\left( \hbar \omega + \mu_{ B } B  \right)^{ 2 } + \hbar^2 v_{ F }^{
2 } \left[ k_{ x }^{ 2 } + \left( k_{ y } + \frac{ e }{ c } B
\langle x \rangle \right)^{2} \right] +  |\sigma| ^{ 2 } + B \kappa
\right \}^2 , \, \label{EqDetA2}
\end{equation}
with $ \kappa = v_{ F }^{ 2 } \hbar ( e / c ) $.

In this expression, by $\sigma$, we mean $\langle 0| \sigma | 0
\rangle $. Analogously $\langle x \rangle$ is the average of the
$x$-coordinate in the lowest energy state of the relativistic Landau
problem \cite{landau}.

The effective potential in the presence of a magnetic field may be
obtained by decomposing the logarithm in two parts and performing
the shift of integration variables $  k_{ y } +  ( e / c)  B \langle
x \rangle\rightarrow k_{ y }$ and $\hbar \omega +\mu_B B \rightarrow
\hbar \omega $, in the first term. Introducing the momentum cutoff
$\Lambda/v_F$, we obtain, up to a constant,

\begin{equation}
V_{\rm eff}\left( |\sigma| , B \right) = \frac{ |\sigma|^{ 2 } }{
\lambda } - \frac{\Lambda }{ \alpha } |\sigma|^{2} + \frac{ 2 }{ 3
\alpha } \left( |\sigma|^{2} + B \kappa \right)^{ \frac{ 3 }{ 2 } }
,\, \label{EqVeff2}
\end{equation}
where $ \alpha = 2 \pi v_{ F }^{ 2 } $.

The divergence can be eliminated as usual by the renormalization
\begin{equation}
\left.
\frac{
\partial^{ 2 } V_{\rm eff}
}{\partial \sigma \partial \sigma^* } \right|_{ |\sigma| = \sigma_{
0 } } \equiv \frac{ 1 }{ \lambda_{ R } } = \frac{ 1 }{ \lambda } -
\frac{ \Lambda }{ \alpha } + \frac{ f( B, \sigma_{ 0 } ) }{
\lambda_{ c } }, \label{EqRenormCondition}
\end{equation}
where $\lambda_{ R }$ is the renormalized coupling parameter, $
\lambda_{ c } = 2 \alpha /3 \sigma_{ 0 } $ and
\begin{equation}
f( B, \sigma_{ 0 } ) = \frac{ 1 + \frac{ 2 B \kappa }{ 3 \sigma_{ 0
}^{ 2 }  } }{ \sqrt{  1 + \frac{ B \kappa }{ \sigma_{ 0 }^{ 2 } }  }
} .\, \label{Eqf}
\end{equation}
In the above equations, $ \sigma_{ 0 }$ is an arbitrary scale, the
renormalization point. In the next section, we perform a
renormalization group analysis, which shows that physical quantities
do not depend on  $\sigma_{ 0 }$.

Using (\ref{EqRenormCondition}) in (\ref{EqVeff2}), we obtain the
renormalized effective potential
\begin{equation}
V_{ { \rm eff},R } \left( |\sigma| , B \right) = \frac{ |\sigma|^{ 2
} }{ \lambda_{ R } } - \frac{ f( B, \sigma_{ 0 } ) }{ \lambda_{ c }
} |\sigma|^{2} + \frac{ 2 }{ 3 \alpha } \left( |\sigma|^{2} + B
\kappa \right)^{ \frac{ 3 }{ 2 } } \, \label{EqVeffR}
\end{equation}

We may now determine the $T=0$ phase diagram as a function of the
magnetic field by analyzing the minima of the effective potential
above. For this purpose, we consider the derivatives of $V_{ { \rm
eff},R } $ with respect to $|\sigma|$, namely
\begin{equation}
V'_{ { \rm eff},R } \left( |\sigma| , B \right) = 2 \sigma \left(
\frac{ 1 }{ \lambda_{ R } } - \frac{ f( B, \sigma_{ 0 } ) }{
\lambda_{ c } } + \frac{ 1 }{ \alpha } \sqrt{ | \sigma |^{ 2 } + B
\kappa } \right) \, \label{EqV'effR}
\end{equation}
and
\begin{equation}
V''_{ { \rm eff},R } \left( |\sigma| , B \right) =
 2  \left(
\frac{ 1 }{ \lambda_{ R } } - \frac{ f( B, \sigma_{ 0 } ) }{
\lambda_{ c } } + \frac{ 1 }{ \alpha } \sqrt{ | \sigma |^{ 2 } + B
\kappa } \right)+ \left( \frac{2 | \sigma |^{ 2 }  }{\alpha \sqrt{ |
\sigma |^{ 2 } + B \kappa } } \right) .\, \label{EqV''effR}
\end{equation}

The minima of the effective potential will occur for  the solutions
of $ V'_{ { \rm eff},R } \left( |\sigma| , B \right) = 0 $, $ V''_{
{ \rm eff},R } \left( |\sigma| , B \right) > 0 $. We, henceforth
call these solutions $\Delta $. Observe that $\Delta =|\langle
0|\sigma |0\rangle|$ and is, therefore, a superconducting order
parameter. The peculiarities of the superconducting transition for
$T\neq 0$ in 2D are well-known and have been discussed extensively in the literature
\cite{2dtrans}.

From (\ref{EqV''effR}) we see that a solution $ \Delta =  0 $ exists
for
\begin{equation}
\lambda_{R} <  \lambda_{ c } (B) = \lambda_{ c } \frac{ \sqrt{1+
\tilde{ B }}  }{1+ \frac{3}{2}  \tilde{ B }  - \frac{2}{3} \sqrt{
\tilde{ B }(1+ \tilde{ B }) } } ,\, \label{EqSigma0Condition}
\end{equation}
where $ \tilde{ B } = B ( \kappa / \sigma_{ 0 }^{ 2 } ) $.

A solution with $ \Delta \neq  0 $ will only occur for $ V'_{ { \rm
eff},R } \left( \Delta , B \right) = 0 $ . From (\ref{EqV'effR}), we
find
\begin{equation}
\Delta_0 = \sqrt{ \alpha^{ 2 } \left( \frac{ f( B, \sigma_{ 0 } )}{
\lambda_{c } } - \frac{ 1 }{ \lambda_{ R } } \right)^{ 2 } - B
\kappa } .\, \label{EqSigmaNeq0}
\end{equation}

Since the first term of (\ref{EqV''effR}) vanishes at this solution,
we can readily see that $ V''_{ { \rm eff},R } \left( \Delta_0 , B
\right)> 0 $, provided $\Delta_0 $ is real, namely, for $\lambda_{R}
> \lambda_{ c } (B)$, where $ \lambda_{ c } (B)$ is given by
(\ref{EqSigma0Condition}). We see that for this range of the
renormalized coupling parameter, the superconducting gap
(\ref{EqSigmaNeq0}) is a true minimum of the effective potential.

\begin{figure}[ht]
\centerline {
\includegraphics
[clip,width=0.9\textwidth ,angle=-90 ] {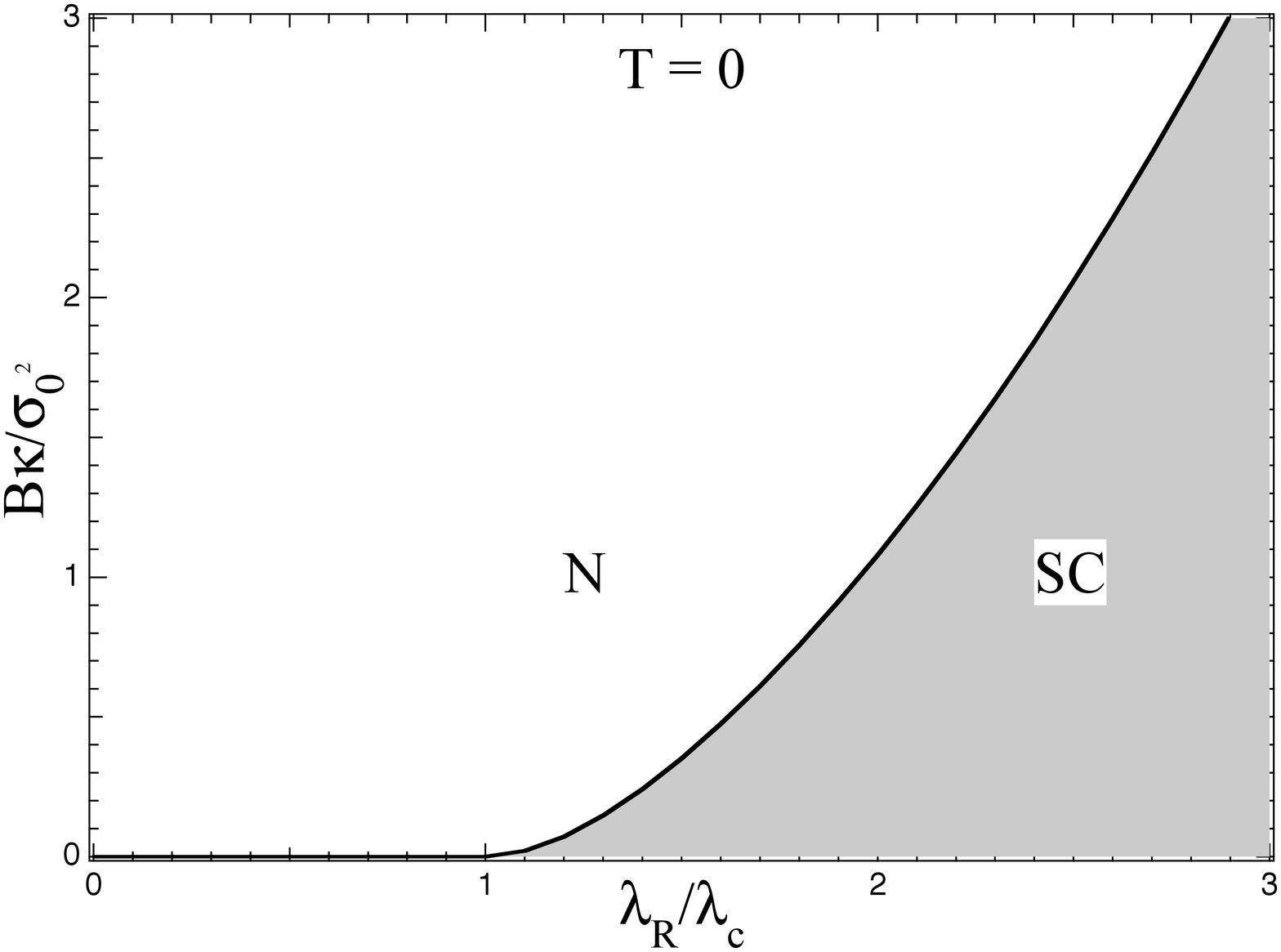} } \caption{The
zero temperature phase diagram of the system.}
\label{FigPhaseDiagramT0}
\end{figure}

The conclusion is that a quantum phase transition connecting a
normal and a superconducting phase occurs at the magnetic field
dependent quantum critical point $ \lambda_{ c } (B)$, given by
(\ref{EqSigma0Condition}). Notice that in the limit $B \rightarrow
0$ both the quantum critical point and the superconducting gap
reduce to the ones found previously \cite{npb1} in the absence of a
magnetic field. Conversely, for each value of the physical coupling
parameter $\lambda_{R}$, there is a critical magnetic field above
which superconductivity is destroyed. This observation allows us to
infer the zero temperature phase diagram of the system, which is
depicted in Fig. \ref{FigPhaseDiagramT0}.

Recent advanced techniques of controlled intercalation of $Cu$ in
$TiSe_2$ enabled the obtainment of the experimental measure of
important physical parameters as a function of doping in this
layered dichalcogenide \cite{tmd}. It would be interesting to
compare this phase diagram with corresponding experimental results
of the critical field as a function of doping in $Cu_x TiSe_2$.

We have shown in \cite{npb1},
that the mean-field phase structure obtained at zero magnetic field is robust
against quantum fluctuations. The same arguments apply here and, therefore, we
reach the same conclusion in the presence of an applied magnetic field.

\section{ Renormalization Group Analysis}\label{Rg}

In order to eliminate the high-momentum divergence, we have
renormalized the theory by performing the subtraction
(\ref{EqRenormCondition}) at the arbitrary scale $\sigma_0$. In this
section, we show that the physical quantities obey renormalization
group equations which show that they are actually independent of
$\sigma_0$.

By inserting the expressions of $\lambda_c$ and of $f( B, \sigma_{ 0
} )$, in (\ref{EqVeffR}), we can show, for instance, that the
effective potential satisfies the renormalization group equation
\begin{equation}
\left( \sigma_{ 0 } \frac{ \partial  }{ \partial \sigma_{ 0 } } +
\beta \frac{ \partial  }{ \partial \lambda_{ R } } \right) \  V_{
\rm eff, R } =0, \, \label{EqRenorm1}
\end{equation}
where the $\beta$-function, defined by
\begin{equation}
\beta \equiv \sigma_{ 0 } \frac{
\partial \lambda_{ R }
}{
\partial \sigma_{ 0 }
} ,\, \label{EqBetaDef}
\end{equation}
is given by
\begin{equation}
\beta = - \frac{ \lambda_{ R }^{ 2 }}{\lambda_c}  \frac{
 \left( 1 + \frac{ 4 }{ 3 }\tilde B
\right) }{ \left( 1 +\tilde B \right)^{ \frac{ 3 }{ 2 } } }  ,\,
\label{EqRenorm2}
\end{equation}
where $ \tilde{ B } = B ( \kappa / \sigma_{ 0 }^{ 2 } ) $.

In analogous fashion, we can show that the superconducting gap
(\ref{EqSigmaNeq0}) also satisfies the renormalization group
equation (\ref{EqRenorm1}), being therefore independent of
$\sigma_0$.

Finally, by integrating the $\beta$-function equation
(\ref{EqBetaDef}), we obtain the result
\begin{equation}
\frac{ 1 }{  \lambda_{ R }  \left( \sigma'_{ 0 }  \right)  } -
\frac{ f \left( \sigma'_{ 0 }, B \right) }{ \lambda_{ c }  \left(
\sigma'_{ 0 }, B \right) } = \frac{ 1 }{  \lambda_{ R }  \left(
\sigma''_{ 0 }  \right)  } - \frac{ f \left( \sigma''_{ 0 }, B
\right) }{ \lambda_{ c }  \left( \sigma''_{ 0 }, B \right) }, \,
\label{EqRenorm3}
\end{equation}
for arbitrary scales $\sigma'_0$ and $\sigma''_0$. This explicitly
shows that the combination
\begin{equation}
\frac{ 1 }{ \lambda_{ R }  \left( \sigma_{ 0 }  \right) } - \frac{ f
\left( \sigma_{ 0 }, B \right) }{ \lambda_{ c }  \left( \sigma_{ 0
}, B \right) }  \, \label{EqRenorm3}
\end{equation}
does not depend on the renormalization scale $\sigma_0$. From
(\ref{EqSigmaNeq0}), then we explicitly see the scale independence
of the superconducting gap $\Delta_0$.

We conclude that the physical properties of the system will be
determined by the physical coupling parameter $\lambda_{\rm R}$,
which should be an experimental input. The value of the
zero-magnetic-field quantum critical point $\lambda_c$ must also be
determined experimentally. Subsequently, the magnetic field
dependence of the quantum critical point may be obtained from
(\ref{EqSigma0Condition}).

\section{Effect of a magnetic field on the superconducting phase transition at $T \neq 0 $}\label{TNeq0}

Let us consider now the effect of an external magnetic field in the
superconducting properties of a quasi-two-dimensional system of
Dirac electrons for $T \neq 0$. In this case, the effective potential
corresponding to (\ref{EqSeff}) is given by
\begin{eqnarray}
V_{\rm eff}\left( |\sigma| , B \right) & = & \frac{ | \sigma |^{ 2 }
}{ \lambda } - 2 \int \frac{ d^{2 } k  }{ \left(  2 \pi \right)^{ 2
} } \sum_{ n = - \infty }^{ \infty } \left\{ \ln \left[ \left( \hbar
\omega_{ n } + \mu_B B \right)^{ 2 } + \hbar^2 v_{ F }^{ 2 } | k |^{
2 } + | \sigma |^{ 2 } + B \kappa \right] \right.
\nonumber \\
& & \hspace{ 3.3cm} \left. - \ln \left[  \hbar  \omega_{ n }^{ 2 } +
v_{ F }^{ 2 } | k |^{ 2 } \right] \right\}, \, \label{EqVeffTNeq0}
\end{eqnarray}
where $\omega_n$ are fermionic Matsubara frequencies, corresponding
to the functional integration over the electron field. Observe that
now we may no longer shift away the magnetic field coupled to the
electron spin. This appears summed to the Matsubara frequencies.

The phase diagram at $T \neq 0$ is obtained from the minima of
(\ref{EqVeffTNeq0}). For the purpose of determining these minima, we
consider the necessary condition $ V'_{\rm eff}\left( |\sigma| , B
\right) = 0 $. Taking the derivative of (\ref{EqVeffTNeq0}) with
respect to $|\sigma|$ and performing the Matsubara sum, we get
\begin{equation}
V'_{\rm eff}\left( |\sigma| , B \right)= 2 |\sigma| \left \{ \frac{
1 }{ \lambda } - \int \frac{ d^{2 } k  }{ \left(  2 \pi \right)^{ 2
} } \ \frac{ 1 }{ E } \left[ \frac{ \sinh\left( \beta E \right) }{
\cosh\left( \beta E \right) + \cosh\left( \beta \ \mu_{ B } B
\right) } \right]\right \} = 0 ,\, \label{EqV'effTNeq00}
\end{equation}
where $ E =  \sqrt{ v_{ F }^{ 2 }k^{ 2 } + | \sigma |^{ 2 } + B
\kappa } $.

Let us look for a superconducting phase. In such a phase, a nonzero
solution of (\ref{EqV'effTNeq00}) (which we call  $\Delta$) is
required. For this, the quantity between round brackets in
(\ref{EqV'effTNeq00}) must vanish. After a change of variables, this
condition leads to the equation for the superconducting gap
$\Delta(T)$:
\begin{equation}
1 = \frac{ \lambda }{ \alpha } \int_{ \sqrt{ \Delta^{ 2 } + B \kappa
} }^{ \Lambda } \ dE \ \frac{ \sinh\left( \beta E \right) }{
\cosh\left( \beta E \right) + \cosh\left( \beta \ \mu_{ B } B
\right) }, \, \label{EqGapEquation}
\end{equation}
where $\Lambda/ v_{ F }$ is the high-momentum cutoff.

The second derivative of the effective potential evaluated at the
solution of (\ref{EqGapEquation}) is given by
\begin{equation}
V''_{\rm eff}\left( \Delta , B \right)=
\frac{2\Delta^2}{\alpha\beta\sqrt{\Delta^2+\kappa B}} \left[
\frac{\sinh\beta\sqrt{\Delta^2+\kappa
B}}{\cosh\beta\sqrt{\Delta^2+\kappa B} + \cosh\beta\mu_B B} \right ]
> 0 .
 \, \label{EqV''effTNeq0}
\end{equation}
This guarantees that the solution of (\ref{EqGapEquation}) is indeed
a minimum of the effective potential.

In order to solve  (\ref{EqGapEquation}), we perform the integration
in $E$. This may be done exactly and the result depends on the
cutoff $\Lambda$. It is a well known fact that the inclusion of a
finite temperature does not change the divergence structure of the
theory. Indeed, the cutoff may be eliminated precisely by the same
renormalization operation (\ref{EqRenormCondition}), which we used
at $T=0$. After this, we obtain the following expression for the
superconducting gap $\Delta(T,B)$, as a function of the temperature
and of the magnetic field,
\begin{equation}
 \Delta^2(T,B)  =
\left \{
k_{ B } T  \cosh^{- 1 }
\left[ \frac{
e^{ \frac{ \sqrt{ \Delta_{ 0 }^{2 } + B \kappa } }{ k_{ B } T }
}  }{ 2 }
- \cosh\left( \frac{ \mu_{ B } B }{ k_{ B } T } \right)
\right]
\right \}^2
- B \kappa  ,
\, \label{EqGapEquationSolution2}
\end{equation}
where $ k_{ B }$ is the Boltzmann's constant and $\Delta_0 \equiv
\Delta(T=0,B)$ is the zero temperature gap, given by
(\ref{EqSigmaNeq0}). It is not difficult to show from the above
equation that indeed, for $T \rightarrow 0$, we have $\Delta(T)
\rightarrow \Delta_0$, as it should.

Now, from the gap expression we may extract the critical temperature
for the superconducting transition, $ T_c $. Indeed, using the fact
that $\Delta(T_c) = 0$ in (\ref{EqGapEquationSolution2}), we get,
\begin{equation}
k_{ B } T_c = \frac{ \sqrt{\Delta_0^2 + B\kappa}}{\ln \left \{2 \left [
 \cosh \left( \frac{\sqrt{B\kappa}}{ k_{ B }T_c}\right)
 +\cosh \left( \frac{\mu_B B}{ k_{ B } T_c}\right) \right ] \right \}} \,
 \label{EqGapEquationSolution3}
\end{equation}
Notice that for $B = 0$, we have $ \Delta_0 (0) / k_{ B } T_{ c }(0)
= 2\ln2 $ in agreement with the result obtained in \cite{npb1}.

We may then re-express the gap as a function of $ T $ and $ T_c $ as
\begin{equation}
 \Delta^2(T,B) =\left \{ k_{ B }T \cosh^{- 1 } \left\{ 2^{(\frac{T_c}{T}-1)}
\left[ \cosh\left(  \frac{\sqrt{\kappa B}}  { k_{ B }  T_c } \right)
+ \cosh\left( \frac{\mu_{ B } B}  {  k_{ B } T_c } \right)
\right]^{T_c/T}
 - \cosh\left( \frac{\mu_{ B } B}  { k_{ B }  T } \right) \right \}\right \}^2 -  B \kappa , \,
\label{EqGapEquationSolution4}
\end{equation}
from which we confirm that the gap vanishes at $T=T_c$.

\begin{figure}[ht]
\centerline {
\includegraphics
[clip,width=0.9\textwidth ,angle=-90 ] {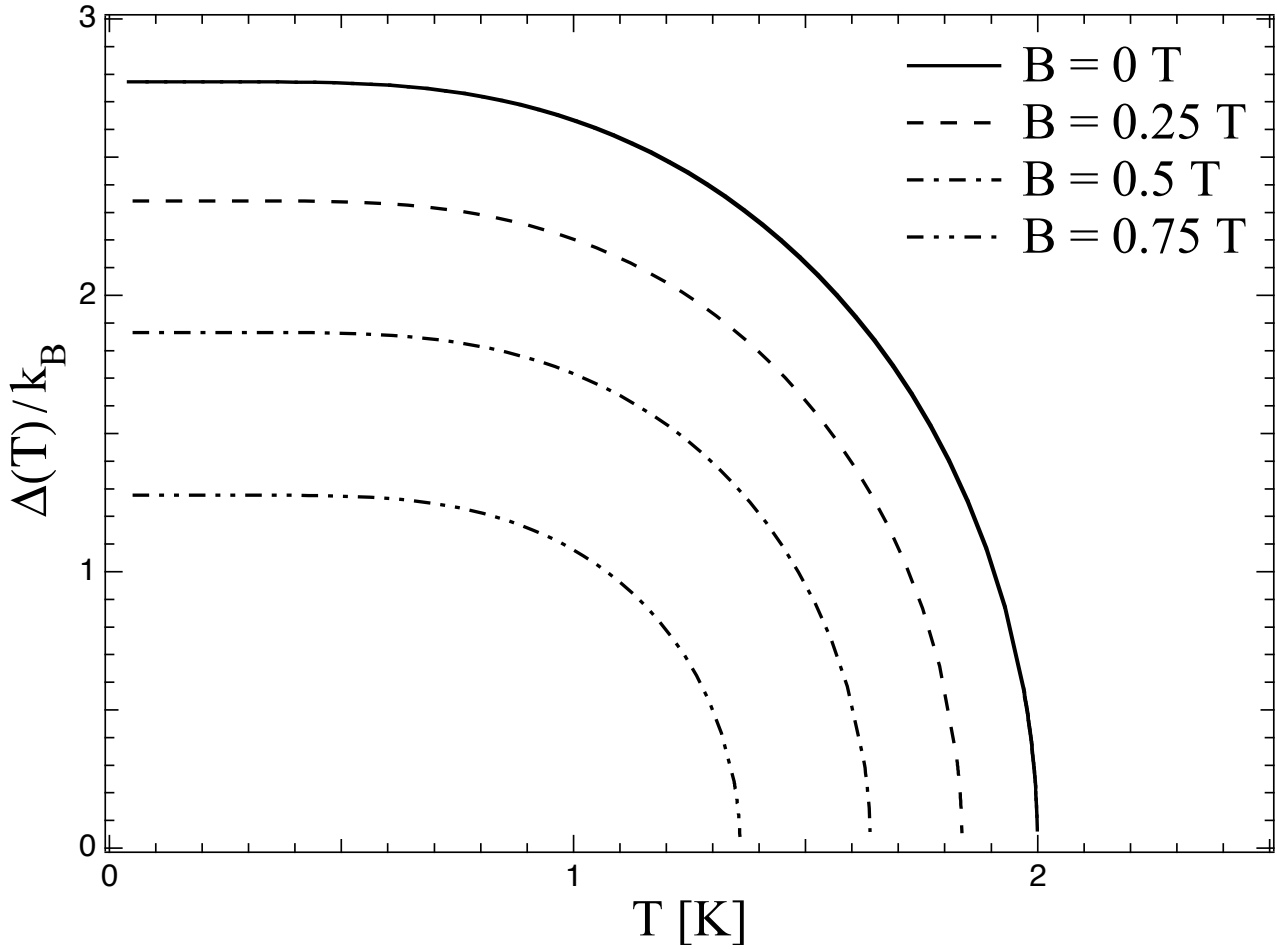} } \caption{The
superconducting gap $ \Delta / k_{ B } $ as a function of the
temperature for several values of the magnetic field. }
\label{FigDeltaxT}
\end{figure}

In order to display the superconducting gap as a function of the
temperature for different values of the magnetic field, we insert
the expression for $ \Delta_{ 0 } $, given by (\ref{EqSigmaNeq0}),
in (\ref{EqGapEquationSolution2}). Our results for $ \lambda_{\rm R}
/ \lambda_{c}  = 2 $ are shown in Fig. \ref{FigDeltaxT} (the
numerical results we present were obtained using  $ v^{ 2 }_{ F } =
1.69 \times 10^{ 8 } $ (m/s)$^{ 2 } $ and $ \sigma_{ 0 } = (8 / 3)
\ln 2 \, k_{B } T_{ c }(0)$ , which yields $ T_{ c }(0) = 2 $ K for
$ \lambda_{ R } / \lambda_{ c } = 2 $ in the absence of magnetic
field).

\begin{figure}[ht]
\centerline {
\includegraphics
[clip,width=0.9\textwidth ,angle=-90 ] {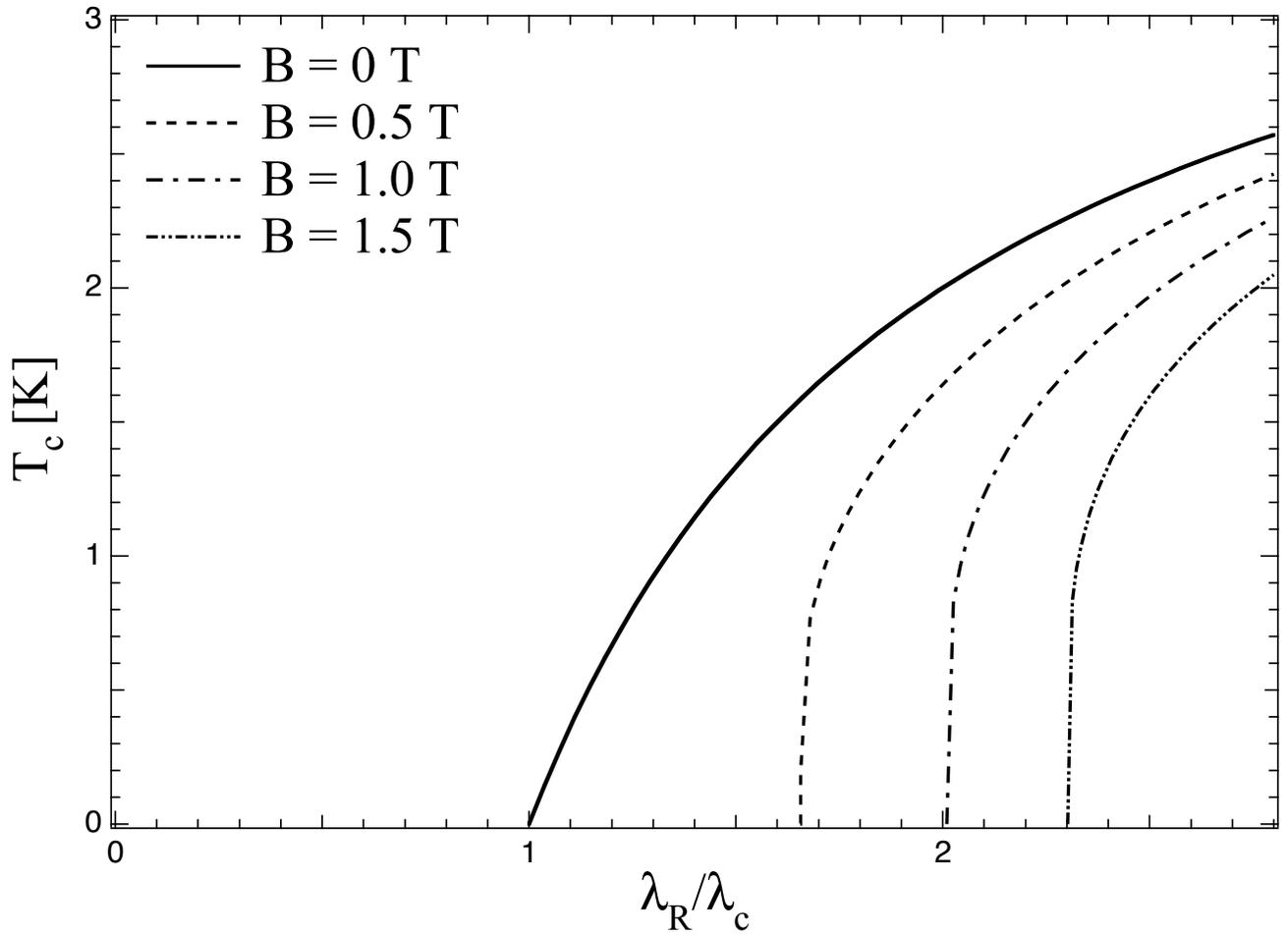} } \caption{$T
\times \lambda_{\rm R} $ phase diagram.The superconducting critical
temperature $ T_{ c } $ as a function of the dimensionless coupling
parameter $\lambda_{\rm R} / \lambda_{c} $ for several values of
magnetic field. } \label{FigTcxX}
\end{figure}

Moreover, by replacing (\ref{EqSigmaNeq0}) in
(\ref{EqGapEquationSolution3}) we may obtain $ T_{ c } $
self-consistently as a function of the coupling parameter. Our
results are depicted in Fig. \ref{FigTcxX}, which is the $T \times
\lambda_{\rm R}$ phase diagram of the system for different values of
the magnetic field. As the magnetic field increases, the quantum
critical point is shifted to the right. Our results suggest that
this behavior may be experimentally verified for the copper-doped
dichalcogenide $Cu_x Ti Se_2$ \cite{tmd}.

\begin{figure}[ht]
\centerline {
\includegraphics
[clip,width=0.9\textwidth ,angle=-90 ] {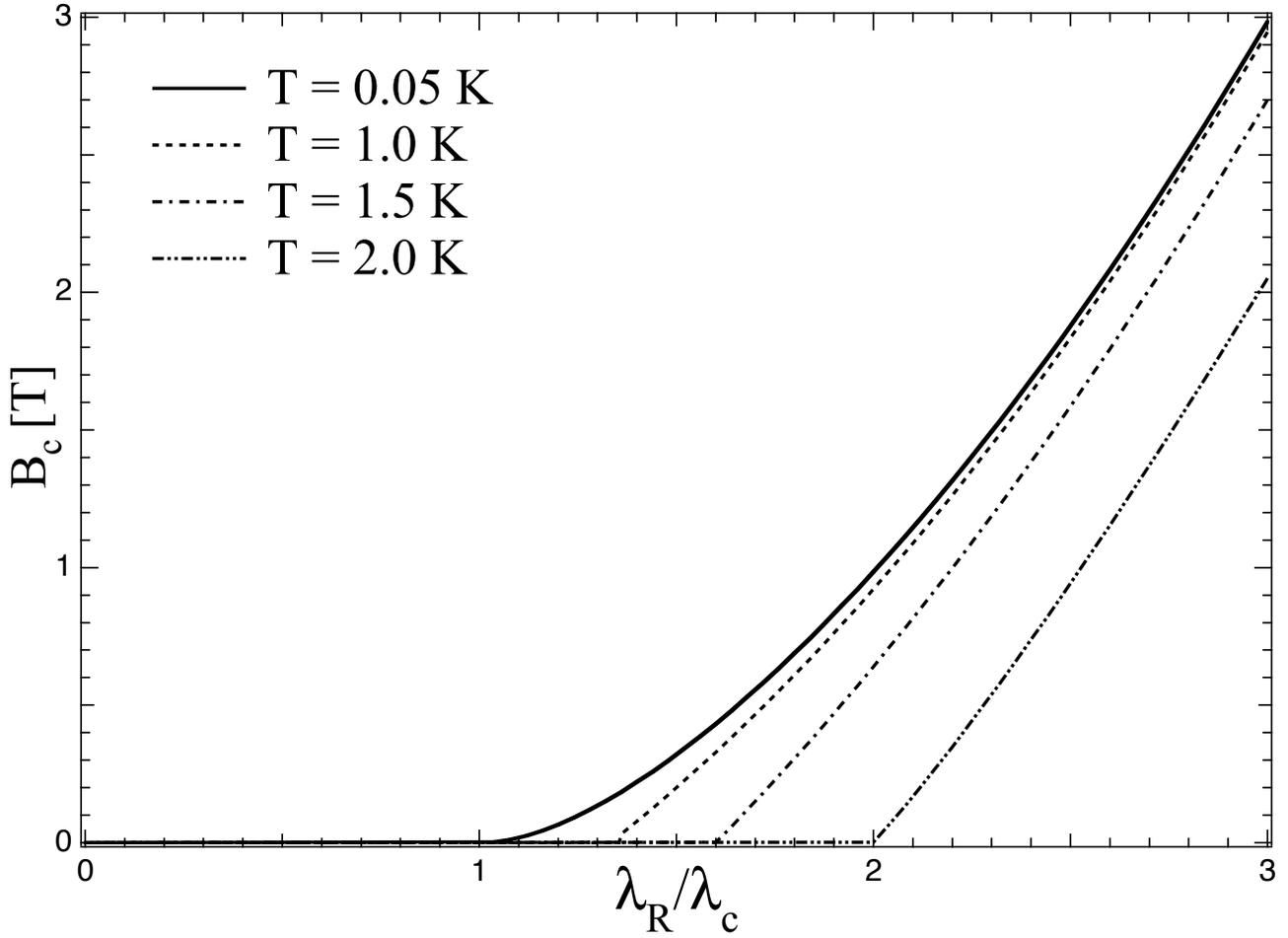} } \caption{$B \times
\lambda_{\rm R}$ phase diagram. The critical magnetic field $ B_{ c
} $ as a function of the dimensionless coupling parameter
$\lambda_{\rm R} / \lambda_{c} $ for several values of the
temperature.  } \label{FigBcxX}
\end{figure}

From (\ref{EqGapEquationSolution3}), we can also obtain the critical
magnetic field as a function of the coupling parameter for different
values of temperature, namely, the $B \times \lambda_{\rm R}$ phase
diagram, which is shown in Fig. \ref{FigBcxX}. We see that for very
low temperatures, it reduces to the same phase diagram displayed in
Fig. \ref{FigPhaseDiagramT0}. However, as the temperature is
increased, the superconducting region, which corresponds to the area
below $ B_{ c } $, becomes smaller.

\begin{figure}[ht]
\centerline {
\includegraphics
[clip,width=0.9\textwidth ,angle=-90 ] {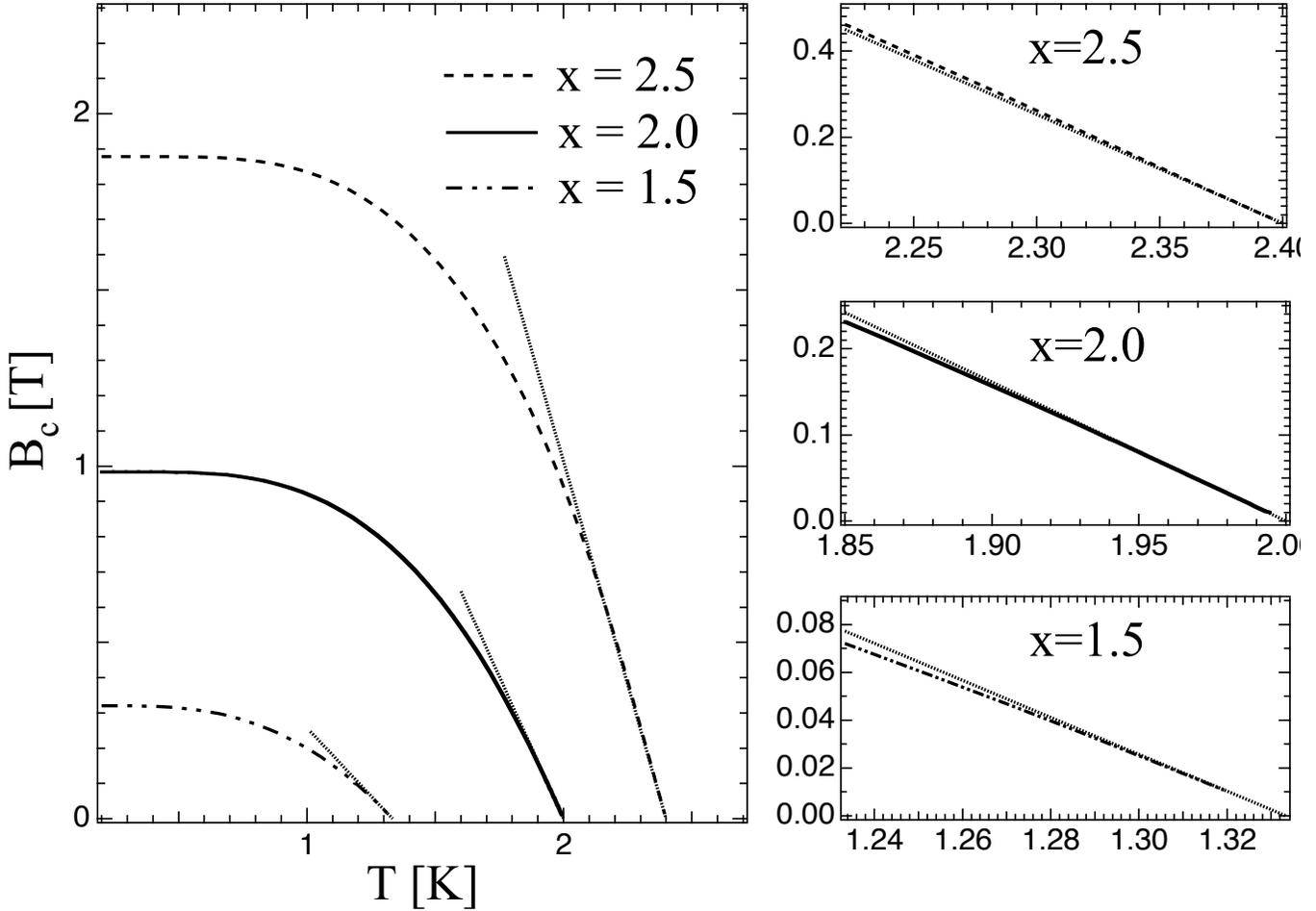} } \caption{$B
\times T $ phase diagram. The critical magnetic field $ B_{ c } $ as
a function of the temperature for several values of the
dimensionless coupling parameter $x \equiv \lambda_{\rm R} /
\lambda_{c} $. The dotted lines in the figure indicate the linear
behavior of $ B $ given by (\ref{EqBcrit}) as $ B \rightarrow 0 $.
 }
\label{FigBcxT}
\end{figure}

From (\ref{EqGapEquationSolution3}), we may also obtain the $B
\times T$ phase diagram for the quasi-two-dimensional
superconducting Dirac electronic system. This is represented in Fig.
\ref{FigBcxT}. Particularly interesting is the linear behavior of
the critical magnetic field for $B \gtrsim 0$. We may derive
explicitly from (\ref{EqGapEquationSolution3}) the following
expression for the critical field, in the small $B$ region:
\begin{equation}
B_c(T) \sim \frac{ 8 \ln 2 k_{ B }^{ 2 } }{ A \kappa } \, { T_{ c
}^{ 2 } ( 0 ) } \left( 1 - \frac{ T }{ T_{ c }( 0 ) } \right) \, ,
\label{EqBcrit}
\end{equation}
where
\begin{equation}
A = 1 - \frac{ 3 }{ 4 \ln 2 } \left( 1 - \frac{ \lambda_{ c } }{
\lambda_{ R } } \right)\, , \label{EqA}
\end{equation}
and
\begin{equation}
k_{ B } T_{ c } ( 0 ) = \frac{ 3 \sigma_{ 0 } }{ 4 \ln 2 } \left( 1
- \frac{ \lambda_{ c } }{ \lambda_{ R } } \right) \, . \label{EqA}
\end{equation}

$B_c(T) $ exhibits a linear behavior of the critical field, which is
indicated by the dotted lines for different values of the
dimensionless parameter $ x \equiv \lambda_{ R } / \lambda_{ c } $
in Fig. \ref{FigBcxT} (this is actually valid for $\lambda_{ R } <
13 \lambda_{ c } $, when $A$ is positive ). This differs from the
quadratic behavior predicted by BCS theory.

A linear decay of the critical field with the temperature similar to
the one obtained here has been experimentally observed in
intercalated graphite compounds \cite{intgraf} and also in the
copper-doped dichalcogenide $Cu_x Ti Se_2$  \cite{tmd}. Since
graphene and also the transition metal dichalcogenides are
well-known to possess Dirac electrons in their spectrum of
excitations, one is naturally led to wonder whether the presence of
such electrons could explain such a behavior of the critical field.

It is interesting to observe that in expressions
(\ref{EqGapEquationSolution2}), (\ref{EqGapEquationSolution3}) and
(\ref{EqGapEquationSolution4}) for the gap and $T_c$, we can trace
back the contributions from the spin and orbital couplings of the
external magnetic field. These are given, respectively by  the
$\mu_{ B }$ and $\kappa$ proportional terms.

\section{Conclusion}\label{Conc}

Starting from the explicit expression for the superconducting gap as a function
of temperature, magnetic field and of the effective superconducting coupling parameter,
we have derived three phase diagrams for a superconducting quasi-two-dimensional system of Dirac electrons.

The $T\times \lambda_{\rm R}$ diagram displays a transition temperature with the same qualitative behavior as the critical temperature
$T_c$ (as a function of doping) observed in high-$T_c$ cuprates in the underdoped region.
It is conceivable that the effective coupling parameter $\lambda_{\rm R}$, controlling the magnitude of the
superconducting interaction, could be effectively determined by the amount of doping. In this case a direct
comparison between our curve and the $T_c \times \text{{\it doping}}$ experimental results for the cuprates would be possible.
Since the coupling magnitude should increase with doping, the curve of $T_c$ as a function of doping would have the
same qualitative form as a function of the coupling $\lambda_{\rm R}$, wich by its turn would qualitatively agree
with ours. The qualitative agreement with the cuprates data
might be
an indication that Dirac electrons play an important role in the mechanism of high-$T_c$ superconductivity.
It would be interesting to investigate the magnetic field dependence of
the superconducting dome in cuprates and especially of the $T=0$ critical doping determining the onset of superconductivity.

The $B \times T$ phase diagram, by its turn shows a linear decay of
the critical field as a function of temperature, for small fields,
which differs from the corresponding quadractic behavior predicted
by BCS theory. It is quite interesting that two materials that are
supposed to have Dirac electrons as elementary excitations, namely,
the copper-doped dichalcogenide $Cu_x Ti Se_2$ \cite{tmd} and
intercalated graphite \cite{intgraf}, both present the same type of
linear behavior of the upper critical field as a function of
temperature, in the small field regime. This might indicate that
Dirac electrons are the common cause of this behavior in both
materials. We intend to explore this point more profoundly in a next
publication.

Finally, the $B \times \lambda_{\rm R}$ phase diagram presents a
quantum critical line, corresponding to a magnetic field dependent
quantum critical parameter $\lambda_c(B)$, connecting the normal and
superconducting phases. Again, we may assume a direct connection
between the coupling parameter $\lambda_{\rm R}$ and the doping
parameter, for instance, in the case of $Cu_x Ti Se_2$. In this
case, we would actually have in Fig. 4 a phase diagram $B \times x$,
where $x$ is the doping parameter. It would be extremely interesting
to have experimental curves of the critical magnetic field as a
function of doping in dichalcogenides such as $Cu_x Ti Se_2$, in
order to compare with our theoretical results.

\section{Acknowledgements}

We would like to thank A.H.Castro Neto and B. Uchoa, for very
helpful comments and conversations. This work has been supported in
part by CNPq and FAPERJ. ECM has been partially supported by CNPq.
LHCMN has been supported by CNPq.

\bibliography{apssamp}


\newpage

\end{document}